\documentclass[floatfix,aps,twocolumn,prl,superscriptaddress]{revtex4}
\usepackage{amsmath}
\usepackage{amssymb}
\usepackage{graphicx}
\usepackage{dcolumn}
\usepackage{natbib}
\usepackage{bm}
\usepackage{epstopdf}
\begin{document}
\title{Towards a two-dimensional superconducting state of La$_{2-x}$Sr$_{x}$CuO$_{2}$ in a moderate external magnetic field}
\author{A. A. Schafgans}
\email{aschafgans@physics.ucsd.edu}
\affiliation{Department of Physics, University of California, San Diego, La Jolla, California 92093, USA}
\author{A. D. LaForge}
\affiliation{Department of Physics, University of California, San Diego, La Jolla, California 92093, USA}
\author{S. V. Dordevic}
\affiliation{Department of Physics, University of Arkon, Akron, Ohio 44325, USA}
\author{M. M. Qazilbash}
\affiliation{Department of Physics, University of California, San Diego, La Jolla, California 92093, USA}
\author{W. J. Padilla}
\affiliation{Department of Physics, University of California, San Diego, La Jolla, California 92093, USA}
\author{K. S. Burch}
\affiliation{Department of Physics, University of California, San Diego, La Jolla, California 92093, USA}
\author{Z. Q. Li}
\affiliation{Department of Physics, University of California, San Diego, La Jolla, California 92093, USA}
\author{Seiki Komiya}
\affiliation{Central Research Institute of the Electric Power Industry, Yokosuka, Kanagawa 240-0196, Japan}
\author{Yoichi Ando}
\affiliation{Institute of Scientific and Industrial Research, Osaka University , Ibaraki, Osaka 567-0047, Japan}
\author{D. N. Basov}
\affiliation{Department of Physics, University of California, San Diego, La Jolla, California 92093, USA}
\date{\today }

\begin{abstract}
We report a novel aspect of the competition and coexistence between magnetism and superconductivity in the high-$T_{c}$ cuprate La$_{2-x}$Sr$_{x}$CuO$_{4}$ (La214). With a modest magnetic field applied $H \parallel c$-axis, we monitored the infrared signature of pair tunneling between the CuO$_2$ planes and discovered the complete suppression of interlayer coupling in a series of underdoped La214 single crystals. We find that the in-plane superconducting properties remain intact, in spite of enhanced magnetism in the planes. 
\end{abstract}

\maketitle
Understanding the interplay between the magnetic and superconducting order parameters in the cuprate high transition temperature ($T_{c}$) materials has presented a substantial experimental and theoretical challenge \cite{Sachdev-Science295-452-2002}. Recent experiments focused on very low-$T_{c}$ Ba-doped and Sr, Nd-co-doped La$_{2}$CuO$_{4}$ and have uncovered that the onset of charge and spin stripelike order leads to the loss of coherence in the superconducting condensate between neighboring CuO$_2$ planes, resulting in a peculiar two-dimensional (2D) superconducing state \cite{Tranquada-PRL99-067001-2007,Tranquada-Unpublished-2008,Tajima-PRL86-500-2001}. In this letter, we show that moderate magnetic fields promote static spin-density wave (SDW) order, which competes with interlayer coherence and leads to a 2D-superconducting state.

Our study focuses on the Josephson plasma resonance (JPR): a collective mode of Cooper pairs oscillating between the CuO$_2$ planes at $T<T_{c}$. The JPR is a signature of bulk 3D superconductivity (SC) in the cuprates \cite{Dordevic-PRL91-167401-2003,Matsuda-PhysC362-64-2001,Bulaevskii-PRB54-7521-1996}, which can be envisioned as a stack of Josephson coupled CuO$_2$ planes, and thus is a sensitive probe of the interlayer phase coherence. The JPR is readily observed using infrared (IR) spectroscopy since it results in the formation of a characteristic plasma edge in far-IR reflectance (Fig. 1a) \cite{Basov-RMP77-721-2005}. By monitoring the JPR in the previously unexplored parameter space of temperature, doping and magnetic field, we are able to investigate the strength of the interlayer phase coherence, allowing us to draw insights into the interplay between magnetism and SC in the La214 system.

The samples in this study were single crystals between 4-6 mm in diameter and 2-3 mm thick with the \emph{ac}-face oriented normal to the incident light. Fabrication and characterization of the samples have been described elsewhere \cite{Komiya-PRB65-214535-2002}. The magneto-optical measurements were performed in an 8 Tesla superconducting magnet at temperatures from 8 K to 295 K and the samples were cooled using He gas exchange \cite{Padilla-RevSciInstrum75-4710-2004}. The field-dependent absolute reflectance $R(\omega)$ was measured over a range of 15 cm$^{-1}$ to 700 cm$^{-1}$. This was combined with zero-field, temperature dependent data ranging from 15 cm$^{-1}$ to 45,000 cm$^{-1}$, to which we applied a Kramers-Kronig analysis in order to extract the optical constants. Low-frequency extrapolations were made using a two-fluid model whereas all high-frequency extrapolations were made assuming a linear regime of $R(\omega)$ eventually decaying as $R(\omega)\propto \omega^{-4}$. 

We start by exploring the JPR in \emph{x}=0.1 La214, well into the underdoped (UD) side of the phase diagram (Fig. 1a, inset). The low energy \emph{c}-axis reflectance $R(\omega)$ in the normal state is insulatorlike, as characterized by a low magnitude and the absence of a metallic plasma edge. There are a series of IR-active phonons in the far-IR region (100-700 cm$^{-1}$), followed by a nearly constant reflectance in the mid-IR range (700-6000 cm$^{-1}$). As the sample is cooled below $T_{c}$, we observe a low-frequency plasma edge develop (the JPR) with a minimum at $\omega=\omega_{p}$, frequencies below which the reflectance approaches unity. With decreasing temperature below $T_{c}$, the JPR moves to higher energies as the superfluid density increases. This is because the JPR is directly related to the superfluid density ($\rho_{s}$) as,
\begin{equation}
\rho_{s}={{c^{2}}\over{\lambda^{2}}}={\omega^{2}_{p}*{\epsilon_{\infty}}}={{4 \pi e^{2} n}\over{m^{*}}}
\end{equation}
where $\lambda$ is the penetration depth, $\omega_{p}$ is the screened plasma frequency and is determined by the number density (\emph{n}) and effective mass (\emph{m*}) of Cooper pairs contributing to the superconducting condensate and the high energy value of the dielectric constant ($\epsilon_{\infty}$).
\begin{figure*}
\centering
\includegraphics[width=7in]{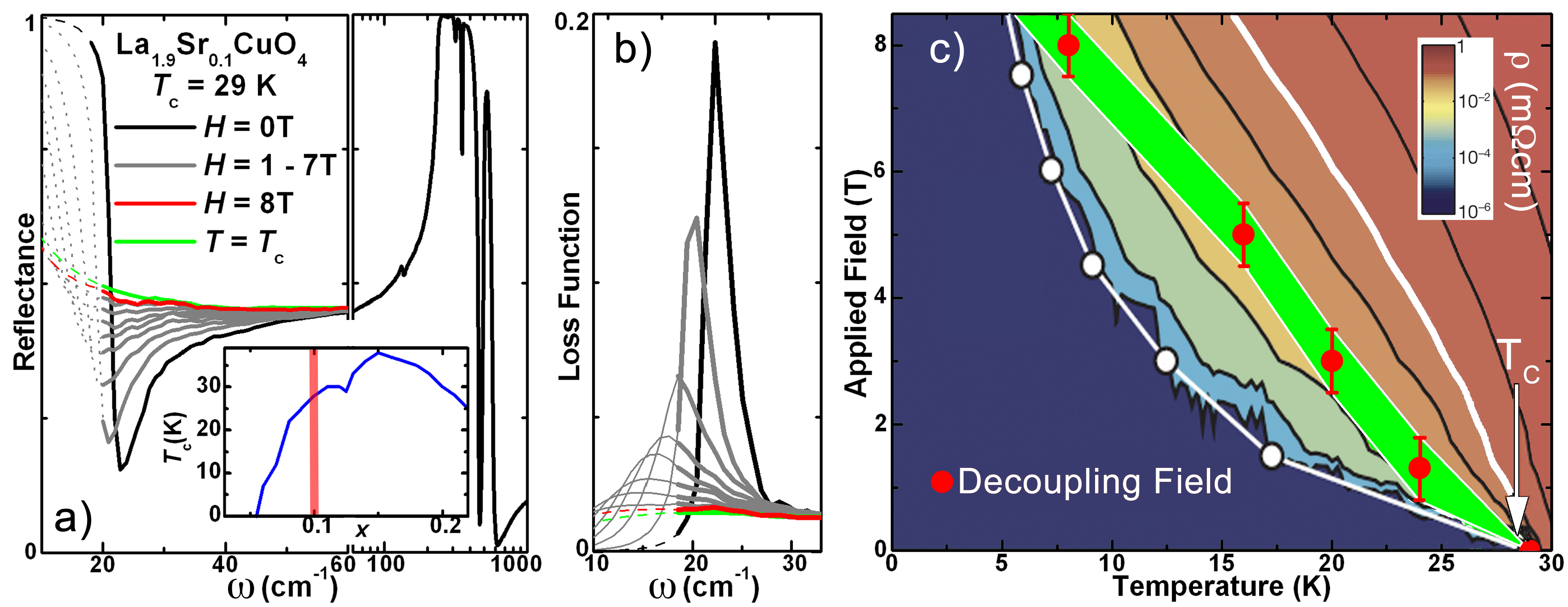}
\caption{(a) Far-infrared reflectance of \emph{x}=0.1 La214 showing the evolution of the JPR at \emph{T}=8 K in magnetic field. By 8 T, the reflectance is restored to the normal state value within the signal to noise of our experiment. The JPR is the only feature in the spectra that is sensitive to the field. Dashed lines are extrapolations. Inset: the superconducting phase diagram of La214 for various Sr content. (b) Loss function at \emph{T}=8 K for \emph{x}=0.1 La214, further described in Fig. 2. (c) Superconducting phase diagram for La214 \emph{x}=0.1, showing the decoupling field $H_{D}(T)$ (red circles) with magneto-resistance data reproduced from Ref.\cite{Lake-Nature415-299-2002}; white circles show the solid-to-liquid vortex phase transition, thick white line (our emphasis) is a constant contour of resistivity near the $T_{c}$ value. Josephson coupling vanishes with increasing field in the green region, the width determined by the uncertainty in $H_{D}$. This region signifies the crossover from 3D to 2D superconductivity.}
\label{Fig. 1}
\end{figure*}

When a magnetic field is applied $H \parallel c$-axis, the JPR decreases in energy, corresponding to a decrease of the \emph{c}-axis superfluid density. Figure 1a shows the field dependence of $R(\omega)$ of the \emph{x}=0.1 sample at \emph{T}=8 K. We find that an applied field of 8 T, well below the upper critical field $H_{c2}(T)$, is sufficient to nearly restore the reflectance to the insulating normal state value. Our observation of the extinction of the JPR in modest fields is an unprecedented and unexpected result \cite{Dordevic-PRB71-054503-2005}. We define the magnetic field that is sufficient to quench the \emph{c}-axis superfluid density below the sensitivity of our measurement to be the decoupling field $H_{D}(T)$.

The temperature dependence of $H_{D}(T)$ is plotted in Fig. 1c for \emph{x}=0.1 (red circles). With increasing temperature, correspondingly smaller fields are needed to quench the JPR. It is instructive to analyze this decoupling line in conjunction with other characteristics of the vortex state; in Fig. 1c we reproduce magneto-resistance data presented by Lake, \emph{et. al.} \cite{Lake-Nature415-299-2002}. White circles represent the solid-to-liquid vortex phase transition and we have emphasized the constant contour of resistivity near the $T_{c}$ value with a thick solid white line. This contour represents an estimate of the resistive critical field $H^{\rho}_{c2}(T)$ \cite{Ando-PRB60-12475-1999}, which can be thought of as the loss of long range phase coherence within the CuO$_2$ planes. The resistive transition to the normal state is gradual. However, $H^{\rho}_{c2}(T)$ should be considered a lower bound on the mean-field pair-breaking field $H_{c2}(T)$ as Nernst effect measurements and specific heat data suggest $H_{c2}(T)$ is much higher and most likely remains temperature independent at low temperatures compared to $T_{c}$ \cite{Wang-PRB73-024510-2006,Wang-arXiv:0803.0638}. Figure 1c shows that the decoupling field is located in the vortex liquid region, well below the loss of long-range superconducting order and the pair-breaking field \cite{Ando-PRB60-12475-1999,Suzuki-PRB60-10500-1999,Wang-arXiv:0803.0638}. In the following, we provide evidence that the decoupling line marks a crossover from 3D SC with prominent Josephson coupling to 2D SC characterized by isolated CuO$_2$ planes. 

In accord with the latter statement, we performed \emph{a}-axis polarized reflection measurements and observed only slight degradation of the in-plane superfluid density in magnetic field by $H_{D}(T)$, within error (Table I) \cite{Savici-PRL95-157001-2005}. The anisotropy of the superfluid density ($\rho_{s}^{a} / \rho_{s}^{c}$) is dramatically enhanced, by at least a factor of 10, due to the depletion of the \emph{c}-axis superfluid in magnetic field. We conclude that superconducting pairing within the CuO$_2$ planes is unharmed by the loss of interlayer coherence. An implication of these results is that an isolated CuO$_{2}$ plane in bulk La214 can maintain high-$T_{c}$ SC.
\begin{table}[ht]
\caption{Superfluid density measured for \emph{x} = 0.1 La214 at \emph{T} = 8 K, with experimentally established upper/lower bounds for in-field suppression}
\centering
\begin{tabular}{c c c c c c}
\hline\hline
\emph{H} (Tesla) & ${\rho_{s}^{a}(H)}\over {\rho_{s}^{a}(0)}$ & $\lambda_{ab}$ (nm) & ${\rho_{s}^{c}(H)}\over {\rho_{s}^{c}(0)}$ & $\lambda_{c} (\mu$m) & ${\rho_{s}^{a}}\over{\rho_{s}^{c}}$ \\ [0.5ex]
\hline
0 & 1 & $406 \pm 9$ & 1 & 12.6 $\pm$ 0.4 & 10$^{3}$ \\
8 & $\geq 0.70$ & $\leq 485$ & $\leq 0.07$ & $\geq$ 48 & $\geq$ 10$^{4}$ \\ [1ex]
\hline
\end{tabular}
\label{table:VWtab}
\end{table}

To illustrate how the in-field behavior changes with doping, Fig. 1b and Fig. 2 present the \emph{c}-axis loss function spectra ($- Im[1/\epsilon(\omega)]$). The loss function quantifies the response of longitudinal modes such as the JPR, which produces a sharp peak centered at $\omega=\omega_{p}$. For the UD samples ($x= 0.1, 0.125$), the peak in the loss function is quenched at low temperatures with only a modest applied magnetic field, while for the near optimally doped (OD) samples ($x= 0.15$ and $0.17$), the suppression is much weaker. These distinctions between the UD and OD crystals are evident from the inspection of the insets to Fig. 2, where we plot the normalized superfluid density ($\rho_{s}(H,T)/\rho_{s0}(T))$ versus applied field. (We used a technique described in Ref. \cite{Dordevic-PRB65-134511-2002} to determine the extrapolation independent superfluid density from the imaginary part of the optical conductivity.) The functional form of $\rho_{s}(H,T)/\rho_{s0}(T)$ is markedly different between the two doping regimes: the UD behavior is sub-linear whereas the OD behavior is entirely linear.
\begin{figure*}
\centering
\includegraphics[width=7in]{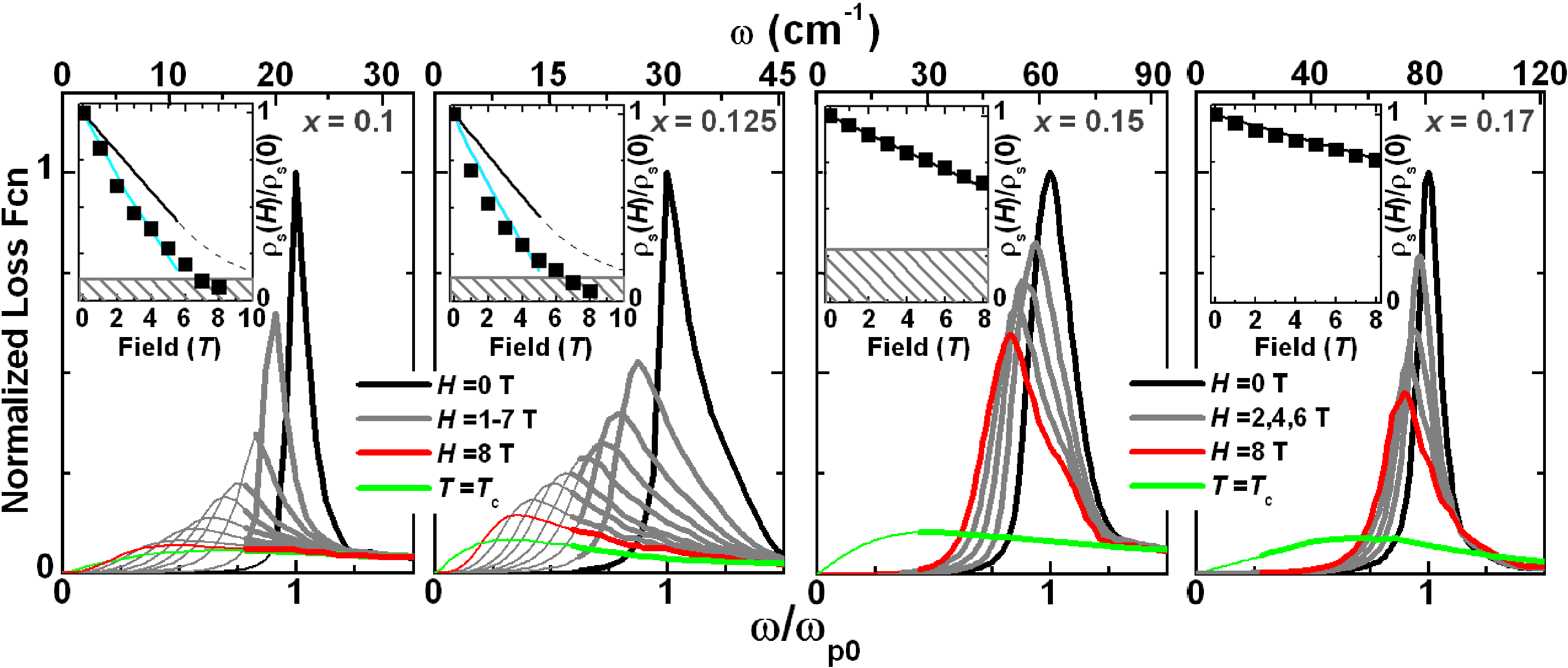}
\caption{Field dependence of the normalized loss function ($-Im[1/\epsilon(\omega)]/-Im[1/\epsilon(\omega_{p0})]$) for the four dopings studied at \emph{T}=8 K. Thick lines are range of extrapolation independent results. Insets: Normalized superfluid density $\rho_{s}(H,T)/\rho_{s0}(T)$ vs. applied field at \emph{T}=8 K. Solid black lines represent the VW model predictions (eqn. 2), and become a dashed guide to the eye at the vortex solid-to-liquid phase transition. Horizontal grey line is the minimum in Josephson coupling allowed by the VW model while values in the grey hashed region are not allowed. In the \emph{x}=0.1 and 0.125 insets, the blue line is of the form in eqn. 5 with A = 1, 1.6 for \emph{x}= 0.1 and 0.125, respectively. The size of the superfluid data points represents the uncertainty due to the form of the low frequency extrapolations. The VW model predicts decoupling fields for \emph{x}=0.1, 0.125, 0.15, 0.17 of $H_{w}(T)$=9.6T, 9.1T, 21 T and 34 T.}
\label{Fig. 2}
\end{figure*}

To explain the suppression of the superfluid density at fields much smaller than the pair-breaking field, we first look to the vortex wandering model (VW) \cite{Koshelev-PRB53-2786-1996,Bulaevskii-PRB54-7521-1996,Bulaevskii-PRB61-R3891-2000}, which describes how the displacement of vortices between neighboring CuO$_2$ planes induces a phase difference in the superconducting order parameter. According to the VW model, the strength of the interlayer coupling energy $E_{j}^{eff}(H,T)$ can be related to the decoupling field due solely to vortex wandering $H_{w}(T)$ \cite{Dulic-PRL86-4660-2001} as
\begin{equation}
{E_{j}^{eff}(H,T)\over{E_{j0}(T)}}=1 - {H\over{H_{w}(T)}}\equiv{v_w(H,T)}
\end{equation}
where $E_{j0}(T)$ is the zero-field coulpling energy and $H_{w}(T)\approx{(E_{j0}(T)+E_{m0}(T))\phi_{0}/U_{p}}$. ($E_{m0}(T)$ is the zero-field magnetic coupling contribution, $\phi_{0}$ is the magnetic flux quanta, and $U_{p}$ is the pinning potential \cite{Koshelev-PRB53-2786-1996}). Equation 2 is valid for fields below the vortex solid-to-liquid phase transition; for higher fields, the predicted suppression of the superfluid loses linearity and becomes $\propto1/H$ \cite{Hwang-PRB59-3896-1999} , moving towards a predicted minimum in interlayer Josephson coupling:
\begin{equation}
{{E_{jmin}\over{E_{j0}}} = {2 \pi E_{j0}\xi^{2}\over{U_{p}}}} 
\end{equation}
(where $\xi$ is the in-plane coherence length). VW cannot drive the interlayer coupling energy to zero and for temperatures $T<<T_{c}$, thermal fluctuations should not play a significant role. For all dopings in our study, $T = 8 K << T_{c}$ satisfies this condition. The normalized superfluid density in magnetic field $\rho_{s}(H,T)/\rho_{s0}(T)$ is a direct probe of $E_{j}^{eff}(H,T)/E_{j0}(T)$ and can be related as ${E_{j}^{eff}(H,T)/{E_{j0}(T)}}={{\rho_{s}(H,T)}/{\rho_{s0}(T)}}$. Therefore, our observation of the interlayer superfluid density is an informative probe of the vortex state.

The experimental results are plotted as the black squares in the insets to Fig. 2 and the VW model prediction is the solid black line in each panel. The VW model works well for \emph{x}=0.15 and \emph{x}=0.17. In UD La214, we find a striking deviation between the suppression of the superfluid density and the VW prediction, differing in two significant ways: first, the superfluid density falls with sub-linear dependence on field; second, the measured superfluid continues below the minimum allowed by VW, trending towards zero. This shows that the VW picture, while being completely adequate for OD samples, is insufficient to describe UD La214.

Data in Fig. 2 call for an additional mechanism that alters the linear law for $\rho_{s}(H,T)$ of the VW model and is capable of completely quenching interlayer Josephson coupling. A distinct property of UD La214 crystals is that an applied field can both stabilize fluctuating magnetism and lead to antiferromagnetic (AF) order extending over macroscopic length scales \cite{Khaykovich-PRB71-220508-2005,Lake-Nature415-299-2002,Sonier-PRB76-064522-2007,Chang-arXiv-07122181v2,Machtoub-PRL94-107009-2005,Savici-PRL95-157001-2005}. Lake, \emph{et.al.}, demonstrated that the field dependence of the normalized ordered spin moment per copper site scales with the applied field as
\begin{equation}
{\mu_{B}^{2}\over{M^{2}}}={H\over{H_{c2}(T)}}\ln{{H_{c2}(T)\over{H}}}
\end{equation}
where \emph{M} is the magnetic moment per copper site and $H_{c2}(T)$ is the temperature dependent upper critical field \cite{Demler-PRL2001}. If interlayer phase decoherence is assisted by static antiferromagnetism, as suggested by experiment \cite{Tajima-PRL86-500-2001} and theory \cite{Berg-Unpub-2008}, then we expect to find a correlation between the field dependence of the in-plane ordered moment and $\rho_{s}(H,T)$. The blue line in the UD insets in figure 2 is of the form
\begin{equation}
{\rho_{s}(H,T)\over{\rho_{s0}(T)}} \propto v_{w}(H,T) - A*{\mu_{B}^{2}\over{M^{2}}}
\end{equation}
This form assumes that interlayer phase decoherence is produced by a concerted action of vortex wandering (eqn. 2) and AF ordering (eqn. 4), with an adjustable fitting parameter A. Equation 5, while being phenomenological, is in remarkably good agreement with our observations and attests to the notion that AF order directly influences interlayer coupling. In UD La214, this happens in such a way that neighboring planes are driven out of phase. Unlike phase decoherence caused by VW, antiferromagnetically driven decoherence increases until the \emph{c}-axis superfluid has been entirely quenched. Thus, field-induced AF order appears to be a viable mechanism responsible for complete decoupling of CuO$_2$ layers and ultimately is the primary cause of the peculiar 2D SC. We stress that field-induced AF order is specific for UD samples and is not found in OD La214, within currently achievable fields.

Our results have direct bearing on reports of the suppression of the JPR in closely related high-$T_{c}$ materials, including Nd-doped La214 and La$_{2-1/8}$Ba${_{1/8}}$CuO$_4$ \cite{Tranquada-PRL99-067001-2007,Tranquada-Unpublished-2008,Tajima-PRL86-500-2001,Homes-PRL96-257002-2006}. These systems reveal the formation of stripelike charge-density wave (CDW) order accompanied by stripelike AF order. In both compounds, not only is the JPR mode frustrated, but the key superconducting characteristics within the CuO$_2$ planes (including $T_{c}$ and the superfluid density) are degraded. Detailed analysis of La$_{2-1/8}$Ba${_{1/8}}$CuO$_4$ reported by Tranquada \emph{et. al.} is suggestive of 2D SC in this compound, albeit with a strongly suppressed transition temperature \cite{Tranquada-Unpublished-2008}. The novelty of the findings reported here is that modest magnetic fields eliminate the JPR in La214 while leaving in-plane SC nearly intact. We stress that in La214, no evidence of field-induced charge order has been identified. Therefore, a correlation between the properties of the JPR and the field-induced magnetic moment conclusively shows that AF spin order alone is the primary competitor of interplane Josephson coupling and is the ultimate reason for 2D SC in this compound. 

The theoretical framework for AF-driven interlayer decoupling is developed in the work of Berg, \emph{et. al.} \cite{Berg-Unpub-2008} with an alternate perspective offered in Ref.\cite{Vojta-arXiv09075202}. An unresolved issue is to experimentally determine the origin of the AF order; is the observed magnetism due to long-range order of vortex cores or due to large patches of AF stripes? Recent realization of scanning tunneling microscopy on La214 is encouraging in the context of resolving this pressing question through direct experiments \cite{Valla-Science-2006}. Finally, we remark that previous reports of Kosterlitz-Thouless (KT) behavior were confined to the vicinity of $T_{c}$ \cite{Li-EPL72-451-2005,Corson-Nature398-221-1999}, whereas our current results extend to $T<<T_{c}$ and may not be KT-like in nature. We present the first observations of a tunable crossover from 3D to 2D SC in a bulk single crystal of a prototypical cuprate. The present work suggests that the phenomenon of 2D SC may be a general characteristic of magnetically ordered cuprates.

We thank S.A. Kivelson, E. Fradkin, and L.N. Bulaevskii for great discussions and acknowledge funding from the NSF and AFOSR MURI. Y. Ando was supported by KAKENHI 19674002 and 20030004.


\begin{thebibliography}{99}

\bibitem{Sachdev-Science295-452-2002} S. Sachdev and S.-C. Zhang, Science \textbf{295}, 452 (2002).

\bibitem{Tranquada-PRL99-067001-2007} Q. Li, \emph{et. al.}, Phys. Rev. Lett. \textbf{99}, 067001 (2007).

\bibitem{Tranquada-Unpublished-2008} J. M. Tranquada, \emph{et. al.}, Phys. Rev. B \textbf{78}, 174529 (2008).

\bibitem{Tajima-PRL86-500-2001} S. Tajima, \emph{et. al.}, Phys. Rev. Lett. \textbf{86}, 500 (2001).

\bibitem{Dordevic-PRL91-167401-2003} S. V. Dordevic, \emph{et. al.}, Phys. Rev. Lett. \textbf{91}, 167401 (2003).

\bibitem{Matsuda-PhysC362-64-2001} Y. Matsuda and M. B. Gaifullin, Physica (Amsterdam) \textbf{362C}, 64 (2001).

\bibitem{Bulaevskii-PRB54-7521-1996} L. N. Bulaevskii, \emph{et. al.}, Phys. Rev. B \textbf{54}, 7521 (1996).

\bibitem{Basov-RMP77-721-2005} D. N. Basov and T. Timusk, Rev. Mod. Phys. \textbf{77}, 721 (2005).

\bibitem{Komiya-PRB65-214535-2002} Seiki Komiya, \emph{et. al.}, Phys. Rev. B \textbf{65}, 214535 (2002).

\bibitem{Padilla-RevSciInstrum75-4710-2004} W. J. Padilla, \emph{et. al.}, Rev. Sci. Instrum. \textbf{75}, 4710 (2001).

\bibitem{Dordevic-PRB71-054503-2005} S. V. Dordevic, \emph{et. al.}, Phys. Rev. B \textbf{71}, 054503 (2005).

\bibitem{Lake-Nature415-299-2002} B. Lake, \emph{et. al.}, Nature (London) \textbf{415}, 299 (2002).

\bibitem{Ando-PRB60-12475-1999} Y. Ando \emph{et al.}, Phys. Rev. B \textbf{60}, 12475 (1999).

\bibitem{Wang-PRB73-024510-2006} Y. Wang, L. Li, and N. P. Ong, \emph{et. at.}, Phys. Rev. B \textbf{73}, 024510 (2006).

\bibitem{Wang-arXiv:0803.0638} Y. Wang and H.-H. Wen, Europhys. Lett. \textbf{81}, 57007 (2008).

\bibitem{Suzuki-PRB60-10500-1999} T. Suzuki, \emph{et. al.}, Phys. Rev. B \textbf{60}, 10500 (1999).

\bibitem{Savici-PRL95-157001-2005} A. T. Savici, \emph{et. al.}, Phys. Rev. Lett. \textbf{95}, 157001 (2005).

\bibitem{Dordevic-PRB65-134511-2002} S. V. Dordevic, \emph{et. al.}, Phys. Rev. B \textbf{65}, 134511 (2002).

\bibitem{Koshelev-PRB53-2786-1996} A. E. Koshelev, L. I. Glazman, and A. I. Larkin \emph{et. al.}, Phys. Rev. B \textbf{53}, 2786 (1996).

\bibitem{Bulaevskii-PRB61-R3891-2000} L. N. Bulaevskii, \emph{et. al.}, Phys. Rev. B \textbf{61}, R3819 (2000).

\bibitem{Dulic-PRL86-4660-2001} D. Dulic, \emph{et. al.}, Phys. Rev. Lett. \textbf{86}, 4660 (2001).

\bibitem{Hwang-PRB59-3896-1999} I.-J. Hwang and D. Stroud, Phys. Rev. B \textbf{59}, 3896 (1999).

\bibitem{Khaykovich-PRB71-220508-2005} B. Khaykovich, \emph{et. al.}, Phys. Rev. B. \textbf{71}, 220508(R) (2005).

\bibitem{Sonier-PRB76-064522-2007} J. E. Sonier, \emph{et. al.}, Phys. Rev. B. \textbf{76}, 064522 (2007).

\bibitem{Chang-arXiv-07122181v2} J. Chang, \emph{et. al.}, Phys. Rev. B \textbf{78}, 104525 (2008).

\bibitem{Machtoub-PRL94-107009-2005} L. H. Machtoub, B. Keimer, and K. Yamada, \emph{et. al.}, Phys. Rev. Lett. \textbf{94}, 107009 (2005).

\bibitem{Demler-PRL2001} E. Demler, S. Sachdev, and Y. Zhang, \emph{et. al.}, Phys. Rev. Lett. \textbf{87}, 067202 (2001).

\bibitem{Berg-Unpub-2008} E. Berg, \emph{et. al.}, Nat. Phys. \textbf{5}, 830 (2009).

\bibitem{Homes-PRL96-257002-2006} C. C. Homes, \emph{et. al.}, Phys. Rev. Lett. \textbf{96}, 257002 (2006).

\bibitem{Vojta-arXiv09075202} A. Wollny and M. Vojta, Phys. Rev. B \textbf{80}, 132504 (2009).

\bibitem{Valla-Science-2006} T. Valla, \emph{et. al.}, Science \textbf{314}, 1914 (2006).

\bibitem{Li-EPL72-451-2005} Lu Li, \emph{et. al.}, Europhys. Lett. \textbf{72}, 451 (2005).

\bibitem{Corson-Nature398-221-1999} J. Corson, \emph{et. al.}, Nature \textbf{398}, 221 (1999).

\end{thebibliography}
\end{document}